\documentclass[%
 reprint,
 amsmath,amssymb,
 aps,
]{revtex4-2}

\usepackage{graphicx}
\usepackage{dcolumn}
\usepackage{bm}
\newcommand\be{\begin{equation}}
\newcommand\ba{\begin{eqnarray}}
\newcommand\ee{\end{equation}}
\newcommand\ea{\end{eqnarray}}

\renewcommand\a{\alpha}
\renewcommand\b{\beta}
\renewcommand\d{\delta}

\renewcommand\l{\lambda}

\renewcommand\t{\tau}

\newcommand\g{\gamma}

\newcommand\m{\mu}

\newcommand\s{\sigma}

\newcommand\pt{\partial}

\newcommand{\non}{\nonumber\\}

\begin{document}

\preprint{APS/123-QED}

\title{Chiral kinetic theory of anomalous transport induced by torsion}

\author{Lan-Lan~Gao}
\affiliation{Physics Department and Center for Particle Physics and
             Field Theory, Fudan University, Shanghai 200438, China}
\affiliation{Department of Physics and Astronomy, Stony Brook University, Stony Brook, New York 11794-3800, USA}            

\author{Sahal~Kaushik}
\affiliation{Department of Physics and Astronomy, Stony Brook University, Stony Brook, New York 11794-3800, USA}

\author{Dmitri~E. Kharzeev}
\affiliation{Center for Nuclear Theory, Department of Physics and Astronomy, Stony Brook University, Stony Brook, New York 11794-3800, USA}
\affiliation{Department of Physics, Brookhaven National Laboratory, Upton, New York 11973-5000, USA}
\affiliation{RIKEN-BNL Research Center, Brookhaven National Laboratory, Upton, New York 11973-5000, USA}

\author{ Evan~John~Philip}
\affiliation{Computational Science Initiative, Brookhaven National Laboratory, Upton, New York 11973-5000, USA}
\affiliation{Department of Physics and Astronomy, Stony Brook University, Stony Brook, New York 11794-3800, USA}

\date{\today}

\begin{abstract}
In Weyl semimetals subjected to torsion, there are two different kinds of chirality: i) the (coordinate-space) shape of the twisted crystal is chiral, and ii) the momentum space contains chiral quasi-particles.  
Here we construct a general kinetic theory of anomalous transport  using the {\it phase space} (coordinate and momentum spaces combined) Berry curvature induced by torsion in Weyl systems. We describe how torsion generates the chiral chemical potential, and thus leads to the Chiral Magnetic Effect (CME) in the presence of a background magnetic field.  We propose to measure the CME current induced by the torsion as a way to detect the anomalous coupling between the coordinate-space and momentum-space chiralities.
\end{abstract}

\maketitle


\section{\label{sec:level1} Introduction}

The defining property of Dirac and Weyl semimetals is the existence of gapless chiral quasiparticles. The spectral flow induced by external parallel electric and magnetic fields leads to the imbalance between the chemical potentials for right- and left-handed chiral quasiparticles, and thus to non-conservation of chiral charge known as the chiral anomaly \citep{Adler1,Bell:1969ts}. The chiral anomaly induces a number of novel phenomena in chiral materials, including the Chiral Magnetic Effect (CME) \citep{fukushima2008chiral} that manifests itself through the negative longitudinal magnetoresistance \cite{son2013chiral,burkov2014chiral} in Dirac materials such as  $\mathrm{ZrTe_5}$\citep{li2016chiral} and $\mathrm{Na_3Bi}$ \citep{xiong2015evidence} and Weyl materials such as TaAs \citep{huang2015observation}. 
Weyl materials also exhibit chirality-dependent optical effects \citep{chan2017photocurrents,de2017quantized,ma2017direct,gao2020chiral} some of which are driven by the chiral anomaly \cite{ma2015chiral,zhong2016gyrotropic,kaushik2019chiral}.
 \medskip 

 It is interesting to consider the effects of crystal deformation on anomalous transport, since deformations are known to lead to strong "synthetic" gauge fields \cite{guinea2010energy,vozmediano2010gauge,
 cortijo2015elastic};for example, in graphene nanobubbles, the synthetic magnetic field in excess of 300 T has been reported \cite{levy2010strain}.  The geometrical torsional response of Weyl fermions has been addressed in \cite{Zhou:2012ix,PhysRevB.99.155152,PhysRevLett.124.117002,PhysRevResearch.2.033269,laurila2020torsional,PhysRevLett.122.056601,PhysRevB.101.165201,Zubkov_2015,sun2014chiral,Hughes_2013}, 
 including its relation to Neih-Yan anomaly  \cite{nieh1982quantized} and chiral torsional effect \cite{Khaidukov:2018oat}.
\medskip

 A natural question to ask is whether these synthetic gauge fields can be used to source the chiral anomaly, and thus drive the CME and other anomalous phenomena. At first glance, it may appear that deformations should be irrelevant for the CME -- indeed, the CME is captured \cite{Kharzeev:2009fn} by the topological Chern-Simons term $\sim \mu_5 \int \epsilon_{ijk} A^i F^{jk}$ in the effective action (where $\mu_5 = \mu_R - \mu_L$ is the chiral chemical potential, and $A^i$ and $F^{jk}$ are the gauge potential and field strength tensor, respectively) that does not depend on the space-time metric $g_{\mu\nu}$. Therefore, 
if one describes the deformation as a change in an effective metric, the chiral anomaly and thus the CME should seemingly not be affected. 
\medskip

However, this conclusion is premature since a time-dependent, inhomogeneous deformation can induce a change in the momentum space distribution of chiral quasiparticles, e.g. by deviating the chiral chemical potential $\mu_5$ away from its equilibrium value $\mu_5=0$ \cite{cortijo2016strain,pikulin2016chiral}. In this case, the CME current will be induced by the deformation in the presence of a background magnetic field \cite{cortijo2016strain,pikulin2016chiral}. The collective excitations mixing sound and chirality have been considered in \cite{song2019hear,Chernodub:2019lhw,PhysRevLett.124.126602}. 

\medskip

Kinetic theory provides a convenient framework for describing transport phenomena. For chiral fermions, the effect of chiral anomaly has been incorporated \cite{Stephanov:2012ki,Son2012Kinetic} in this theory via the Berry curvature; see  \cite{xiao2010berry} for a review of earlier work on Berry curvature effects on transport. In the resulting chiral kinetic theory, Weyl cones are described as momentum-space monopoles of Berry curvature. In the presence of an external magnetic field, the combined effect of the momentum-space monopoles and magnetic field is the modification of the density matrix in the phase space, resulting in the anomalous Hall effect and the CME  \cite{Stephanov:2012ki,Son2012Kinetic,Chen:2012ca,Dwivedi:2013dea,Basar:2013qia,Basar:2013iaa,Manuel:2013zaa,Chen:2014cla,Gorbar:2016ygi,Kharzeev:2016sut}.
\medskip

In this paper, our goal is to construct a chiral kinetic theory for Weyl materials under a mechanical twist, i.e. Saint-Venant torsion. We focus on a twisted crystal under non-uniform strain, similar to the case studied in \cite{grushin2016inhomogeneous}. We will first show that the effect of dynamical, time-dependent deformations on chiral fermions can be captured by Berry curvature in phase space (coordinate and momentum spaces combined).  This quantity has been introduced and used before in a variety of problems involving chirality and spatially inhomogeneous backgrounds  \cite{sundaram1999wave,volovik2003universe,xiao2005berry,freimuth2013phase}. In our problem, Berry curvature in phase space emerges because the spatially inhomogeneous, time-dependent deformations change the distribution of the fermions both in coordinate and momentum spaces.  We then derive the generalized chiral kinetic equations (\ref{eq:kineq4}) describing the effect of deformations on transport of chiral fermions, both with and without external electromagnetic fields. The corresponding anomaly equation (\ref{eq:kineq3}) has, as a source of chirality, an exterior derivative of the phase-space Berry curvature. Since it measures the charge of the Berry monopole in phase space, we call it the "monopole charge function". 

The physical meaning of the anomaly equation is as follows: the angular twist creates a synthetic magnetic field $B_{eff}$, the time-dependent strain creates a synthetic electric field $E_{eff}$, and they combine to yield a source $E^i_{eff} B^i_{eff}$ for the chiral charge generation in the $n_{th}$ cone. Although the anomalous current induced by strain alone cancels out after summation over all Weyl cone pairs, the strain does provide a total chiral imbalance $\m_5$ -- therefore, in the presence of an external magnetic field, the chiral magnetic current appears. We then use the generalized chiral kinetic theory to evaluate the magnitude of the CME current induced by torsion in the presence of magnetic field. The current has a linear dependence both on the chiral chemical potential and the magnetic field, similarly to the "usual" CME. It is observable, and provides a way to discover the generation of chiral imbalance through synthetic gauge fields arising from the phase space Berry curvature. Throughout the paper, we use as an example a concrete tight-binding Hamiltonian (\ref{tight}); however all of our derivations apply to any Hamiltonian of type $H = {\vec \sigma}\cdot {\vec p}(k, x, t) + \phi($k$, $x$, t)$ without an explicit cut off.


\medskip

Our work focuses only on elastic deformations without the defects, e.g. screw dislocations. 
 Screw dislocation is a topological defect in the crystal structure and hence also in the synthetic gauge field, which acquires a delta-function analogous to the Abrikosov vortex in a type II superconductor.  The elastic torsion we consider deforms the lattice, and therefore the Hamiltonian, continuously, whereas the defects induce singularities, and thus potential sensitivity to the UV cutoff \cite{PhysRevLett.116.166601}.
 
Let us consider a angular twisted Weyl semimetal with time-dependent compression along the axis of twist. The twist is chiral; the compression and twist together generate chirality and produce a chiral imbalance for the Weyl quasiparticles, as we will demonstrate. 
In order to solve this problem within the framework of chiral kinetic theory, we need to do the following. 
First, we need to understand how the elastic deformation in position space affects the Berry curvature, in both position and momentum spaces.
Second, we need to formulate and solve the kinetic equation, and calculate the anomalous current and chiral charge generation.

For the first problem, because the Hamiltonian in momentum space now depends on the spatial coordinates, it is impossible to separate the position space and  momentum space components of Berry curvature. 
Therefore, we will address this problem from the perspective of the phase space path integral; our starting point is a phase space Lagrangian, which we will obtain from the coordinate-dependent Hamiltonian.
To solve the second problem, we start from the definition of the Berry connection in phase space and  derive the corresponding  kinetic equation.


 \medskip

\section{Hamiltonian in deformed Weyl semimetal}
Since many work on Hamiltonian under deformation is derived from tight-binding model\cite{Cortijo_2016}\cite{de2012space}\cite{volovik2014emergent}
\cite{Shapourian_2015}.We will use a simple model Hamiltonian of an anisotropic Weyl semimetal proposed in \cite{mccormick2017minimal} as an example. This model has a simple tetragonal lattice with lattice constants $a$ and $b$. 
The tight-binding Hamiltonian of the model is given by
\begin{equation}\label{tight}
\begin{split}
H= &-\sum_i\,(c_{i-1x'}^+c_{i}+c_{i+1x'}^+c_{i})\, t_1\s_x\\&+2cos(k_{0} a)\cdot \sum_ic_i^+c_i\, t_1\s_x\\\ &+i\sum_i\,(c_{i+1\perp'}^+c_{i}-c_{i-1\perp'}^+c_{i})\,t_2\s_\perp\\ &+ \sum_i (c_{i+1\perp'}^+c_{i}+c_{i-1\perp'}^+c_{i})t_3
\end{split}
\end{equation}
where $a$ is the lattice spacing in x-direction, and $c^+$ and $c$ are creation and annihilation operators. The parameters $t_1$, $t_2$ and $t_3$ represent the strength of tight-binding interaction, and $k_{0}$ is the location of Weyl points in momentum space.  The $\perp$ denotes the directions perpendicular to the specific $x$-axis, so it corresponds to the indices in $y-z$ plane, which can be taken as $\hat{y}$ or $\hat{z}$. For example $c_{i+1\perp'}\sigma_\perp$ means $c_{i+1y'}\sigma_y+c_{i+1z'}\sigma_z$.

This model possesses a fourfold rotation symmetry around the $x$-axis, and inversion symmetry, but not the time-reversal symmetry. The deformation can be described by the following parameter change:
\begin{multline}
    (c_{i-1x'}^+ c_i + c_{i+1x'}^+c_i) t_1\s_x \rightarrow \\(c_{i-1x'}^+ c_i + c_{i+1x'}^+c_i) (t_1\s_x-\beta_1u_{xx}\s_x) +\\ i(c_{i-1x'}^+c_i - c_{i+1x'}^+c_i)\beta_2 u_{x\perp}\cdot \s_{\perp},
\end{multline}
\begin{equation}
t_2\s_\perp\rightarrow t_2\s_\perp -\beta_3 u_{\perp\perp}\cdot \s_{\perp},
\end{equation}
where $\beta$ is the anisotropic Gruneisen parameter and $u_{ab}$ is the strain tensor.
We set the twist to be along the $x$-axis,
$u_{x\perp}=\gamma\varepsilon_{ij} r_j$ ; $\g$ describes the twist angle gradient, and $\varepsilon_{ij}$ is the rank two antisymmetric tensor in $y-z$ plane. We also apply a time-dependent compression $\lambda = - u_{xx}$.

With these substitutions, the strain-modified  Hamiltonian in momentum space becomes:

\begin{equation}
\label{eq:Ham} 
\begin{split}
H= &-2t_1(\cos(k'_x a)-\cos(k_{0x} a))\s_x \\ &-2t_2\sin(k'_y b)\s_y
- 2t_2\sin(k'_z b)\s_z\\&-2\beta_2\gamma y \sin(k'_x a)\s_z+2\beta_2\gamma z \sin(k'_x a)\s_y\\ &+ 2 \b_1 \cos(k'_xa) \lambda\s_x + 2 [\cos(k'_y b) + \cos(k'_z b)] t_3 ,
\end{split}
\end{equation}
where the primed symbols denote the local lattice vector direction. The transformation of direction from global vierbein to the local one is given by
\begin{equation}
  \label{eq:metric1}
\begin{pmatrix}
   d x\\
d y\\
d z\\
    \end{pmatrix} =
    \begin{pmatrix}
    1-\lambda(t)& 0 & 0\\
0 &1&-\g x'\\
0& \g x'&1\\
    \end{pmatrix}\cdot \begin{pmatrix}
   d x'\\
dy'\\
d z'\\
    \end{pmatrix};
\end{equation}
we will denote the transformation matrix as $M_i^j=\frac{dx^j}{dx'^i}$.


Expanding the effective Hamiltonian in the vicinity of the Weyl point at momentum $K_i$,
we get a familiar form as
\begin{equation}
\begin{split}
\label{eq:Ha}
&H_{eff}={e'}_a^i\s_a (k'_i-K'_i)+{W'}^i(k'_i-K'_i)+\mathcal E\\
&\qquad=e_a^i\s_a(k_i-K_i)+W^i(k_i-K_i)+\mathcal E,
\end{split}
\end{equation}
where the local Weyl points $K'$, dreibein $e'$ ,tilt vector $W'$ and energy of Weyl points $\mathcal E$ can be derived from equation~\eqref{eq:Ham}.
 The corresponding dreibein and Weyl points in global coordinates, which are defined to be consistent with the untwisted lattice, are given by $K_i=(M^{-1})_i^j K'_j$, $e_a^i={e'}_a^j M^i_j$ and $W^i={W'}^{j}M_j^i$.
 
In the next section, we will prove that the elastic chiral anomaly is independent of $e_a^i$ and $W^i$, so there is no need to give an explicit form of these quantities here. All we need is the elastic gauge field that can be read off the form of the effective Hamiltonian.
There are eight Weyl cones. In the lowest order:
\begin{equation}
\begin{split}
K_x &= s_x [k_{0x} + \frac{\b_1}{t_1 a}\cot(k_{0x}a)\lambda]\\
K_y &= s_y\frac{\beta_2 \gamma  z'}{t_2 b}s_x \sin(k_{0x} a) +\Theta(-s_y) \frac{\pi}{b}\\
K_z &=-s_z\frac{\beta_2 \gamma y'}{t_2 b}s_x \sin(k_{0x} a) +\Theta(-s_z))\frac{\pi}{b}
 \end{split}
\end{equation}
where $s_x = sgn(\sin(K_x a))$ and $s_{y,z} = sgn(\cos(K_{y,z} a))$, the chirality $\chi$ is $s_x s_y s_z$.
The velocity of fermions in the $x$ direction is $v^x = 2at_1 cos(k_{0x}a)$ and in the $y,z$ direction is $v^{y,z} = 2t_2 b$.

\medskip

\section{ Kinetic equations for the Hamiltonian $\sigma\cdot p(k,x,t)+\phi(k,x,t)$.}
In the above analysis, we find that the elastic deformation affects the dreibein, tilt vector and position of Weyl points, but the Hamiltonian always maintains the form $\sigma\cdot p(k,x,t)+\phi(k,x,t)$. This Hamiltonian can be analyzed by chiral kinetic theory from a path integral perspective \cite{Stephanov:2012ki}, where the action is given by
\begin{equation}
 \label{eq:I}
I=\int_{t_i}^{t_f}(k\cdot \dot{x}-\epsilon(p)-a_k\cdot \dot{k}-a_x\cdot \dot{x}-a_t )dt ;
\end{equation}
$\epsilon = \pm|p|+\phi $ is the eigenvalue of energy.  

Generalizing the Berry connection and curvature in momentum space \cite{Stephanov:2012ki}, the Berry connection and Berry curvature in phase space can be defined as:
$a_\a=i\langle u|\frac{\pt p_i}{\pt \a}\pt_p|u\rangle =\frac{\pt p_i}{\pt \a}a^p_i $, $\Omega_{\a\b}=\pt_\b a_\a-\pt_\a a_\b=\frac{1}{2}\pt_\a p_m \pt_\b p_l\epsilon_{mln}\Omega_n^p$, where $\Omega_n^=\frac{\hat{p}_n}{2|p|^2}$  Below, we use the Greek characters as the indices in phase space.

For an arbitrary Hamiltonian, the kinetic equations corresponding to ~\eqref{eq:I} are derived in \cite{Hayata:2016wgy}; see equations (C3,C8,C9) in that paper. For our Hamiltonian, the Berry curvatures satisfy the following identity:
\begin{equation}
 \label{eq:Weq-full}
 \begin{split}
  & \Omega_{\a\b}\Omega_{\g\s}+\Omega_{\a\g}\Omega_{\s\b}+\Omega_{\a\s}\Omega_{\g\b}=0.
 \end{split}
\end{equation}

Using this identity, the kinetic equations can be brought to the form
\begin{multline}
\label{eq:kineq4} 
\sqrt{G}\dot{x}^i=-\Omega_{k_i t}+(\delta^i_j(1+\Omega_{k_lx^l})-\Omega_{k_ix^j})\pt_{k_j}\epsilon\\+\Omega_{k_ik_j}\pt_{x^j}\epsilon,
\end{multline}
\begin{multline}
\label{eq:kineq2} 
\sqrt{G}\dot{k}_i=\Omega_{x^i t}-(\delta_i^j(1+\Omega_{k_lx^k})-\Omega_{k_ix^j})\pt_{x^j}\epsilon\\-\Omega_{x^ix^j}\pt_{k_j}\epsilon.
\end{multline}
where $\sqrt{G}=1+\Omega_{k_ix^i}$ describes the modification of the phase space density.

The corresponding anomaly equation can be written as:
\begin{multline}
\label{eq:kineq3} 
\frac{\pt\sqrt{G}}{\pt t}+\frac{\pt\sqrt{G}\dot{x}^i}{\pt x^i}+\frac{\pt\sqrt{G}\dot{k}_i}{\pt k_i}=\Theta_{k_i x^i t}+\Theta_{k_j x^jx^i}\pt_{k_i}\epsilon\\+\Theta_{x^jk_jk_i}\pt_{x^i}\epsilon.
\end{multline}
where 
\begin{equation}\label{mcf}
\Theta_{\a\b\g}=\pt_\a\Omega_{\b\g}+\pt_\b\Omega_{\g\a}+\pt_\g\Omega_{\a\b}
\end{equation}
 is the exterior derivative of the Berry curvature; we will call this quantity the monopole charge function, as it measures the charge of the Berry monopole in phase space. 
 
It is easy to prove that if the Berry connections are all continuous analytical functions, the equation (12) becomes a classical Liouville equation, because $\Theta_{\a\b\g}=0$. However the Berry curvature possesses a singularity at the point $p=0$, where two degenerate bands cross. This is the source for the chiral anomaly.

The chiral anomaly is linked to the topology
of fields in the system; To investigate the gravitational topology and gauge topology, we need to understand the source of $e^i_a$ and $K_i$.

As we showed in the first section, to form the dreibein $e_a^i$ we need the following: 1) the elastic deformation of the lattice that changes the dispersion relation; and 2) the frame transformation matrix $M_i^j$ induced by the geometric structure.  The gauge-analogy effect is affected only by the existence of the  Weyl points $K_i$, and is independent of the dreibein $(e)_a^i$ and the tilt $W^i$ that describe the shape of the cones. The gravitational "Nieh-yan" anomaly effect is determined by nontrival structure (for example, dislocation) and the momentum cut-off. The structure is absent in our continuous deformation model. So the final result for the anomaly is cut-off-independent. 
In other words, the chiral anomaly in our model just contains the gauge-analogy anomaly which depends on the existence of Weyl cones($K_i$), but not on their detailed shapes, as we will now demonstrate.

\medskip

\section{Consistency check of kinetic equations.}
Let us first check that our equations are consistent with \cite{Stephanov:2012ki} in the absence of deformations. In this case, 
$p_i=k_i-A_i$; we will use the Coulomb gauge.
From (\ref{mcf}), we get for the monopole charge function
\begin{equation}
\Theta_{\a\b\g}=\varepsilon_{mnl}\frac{\pt p_m}{\pt \a}\frac{\pt p_n}{\pt \b}\frac{\pt p_l}{\pt \g}\Theta_p,
\end{equation}
where $\Theta_p=2\pi\d^3(p)$ is the well known Berry monopole in momentum space. Therefore,
\begin{eqnarray}
&\Theta_{k_ix^it}=2\pi(\vec{E}\cdot \vec{B})\d^3(p) ,\\
&\Theta_{k_jx^jx^i}\pt_{k_i}\epsilon+\Theta_{x^jk_jk^i}\pt_{x^i}\epsilon=0 .
\end{eqnarray}
 We thus obtain from (\ref{eq:kineq3}) the same expression for the ``Liouville anomaly" as in \cite{Stephanov:2012ki}:
\begin{equation}
\frac{\pt\sqrt{G}}{\pt t}+\frac{\pt\sqrt{G}\dot{x}_i}{\pt x^i}+\frac{\pt\sqrt{G}\dot{k}_i}{\pt k_i}=2\pi(\vec{E}\cdot \vec{B})\d^3(p) .
\end{equation}
Let us now check the consistency of kinetic equation~\eqref{eq:kineq4}. For the Weyl Hamiltonian $H = \sigma_i k_i$ we get 
\begin{eqnarray}
 &\Omega_{k_it}= -\vec E\times  \frac{\hat{p}}{2|p|^2},\qquad\Omega_{k_lx^l}\pt_{k_i}\epsilon=(\vec B\cdot \vec \Omega)\hat{p},\\
 & -\Omega_{k_ix^j}\pt_{k_j}\epsilon+\Omega_{k_ik_j}\pt_{x^j}\epsilon =(\hat{p}\cdot\vec \Omega)\vec B-(\vec B\cdot \vec\Omega)\hat{p}.
 \end{eqnarray}
The kinetic equation is thus given by
\begin{equation}
\sqrt{G}\dot{x}=\hat{p}+\vec E\times  \frac{\hat{p}}{2|p|^2} +(\hat{p}\cdot\vec \Omega)\vec B ,
\end{equation}
which is also same with the equation derived by Stephanov and Yin \cite{Stephanov:2012ki}. 
\medskip

\section{Derivation of topologically protected CME and chiral anomaly for a general deformed Weyl semimetal }
Let us now apply the kinetic equations \eqref{eq:kineq4}-\eqref{eq:kineq3}
to the Weyl semimetal under torsion. 
We first use the anomaly equation \eqref{eq:kineq3} to identify the quantities responsible for the topological configuration of the system. In our case, substituting  $p_a=e_a^i(x)(k_i-A_i-K_i)$, and $\phi=W^i(x)(k_i-A_i-K_i)+\mathcal E(x)$. 
 into \eqref{eq:kineq3}, we get
\begin{equation}
\begin{split}
\label{eq:theta}
&\Theta_{k_i x^i t}=e_a^i e_b^j\frac{\pt (A_{j}+K_j)}{\pt x_i}e_c^l\frac{\pt (A_{l}+K_l)}{\pt t}\varepsilon_{abc}2\pi\d^3(p)\\
&\qquad\qquad-e_a^i\frac{\pt e_b^j}{\pt x^i}T_j e_c^l\frac{\pt (A_{l}+K_l)}{\pt t}\varepsilon_{abc}2\pi\d^3(p)\\
&=2\pi det(e)(\frac{d(\vec{A}+\vec{K})}{dt}\cdot[\nabla\times(\vec{A}+\vec{K})])\d^3(p)
\end{split}
\end{equation}
where $T_i =e^{-1}_{ij}p_j= k_i - A_i - K_i$.  The second line above vanishes by $p_i\d^3(p)=0$.
 
 The monopole charge functions $\Theta$ only depend on $p$, so \eqref{eq:theta} is tilt independent. 
One can further prove that the last two terms in \eqref{eq:kineq3} are also same:
\begin{equation}
\begin{split}
&\Theta_{k_jx^jx^i}\pt_{k_i}\epsilon+\Theta_{x^jk_jk_i}\pt_{x^i}\epsilon\\
&=2\pi det(e)(\nabla \mathcal E\cdot[\nabla\times(\vec{A}+\vec{K})])\d^3(p)
\end{split}
\end{equation}
Therefore, the anomaly equation for an isolated Weyl cone is:
\begin{multline}
\label{eq:topochar}
\pt_\m j^\m =\int \frac{d^3k}{(2\pi)^3}\left(\frac{\pt\sqrt{G}}{\pt t}+\frac{\pt\sqrt{G}\dot{x}^i}{\pt x^i}+\frac{\pt\sqrt{G}\dot{k}_i}{\pt k_i}\right)f\\
=\int sgn(J)\frac{J^{-1}d^3p}{(2\pi)^3}(\Theta_{k_i x^i t}+\Theta_{k_jx^jx^i}\pt_{k_i}\epsilon\\
+\Theta_{x^jk_jk_i}\pt_{x^i}\epsilon)f\\
=\frac{sgn(J)}{4\pi^2}det(e)^{-1}det(e)\vec{B}_{eff}\cdot \vec{E}_{eff}\\
=\frac{\chi \vec{B}_{eff}\cdot \vec{E}_{eff} }{4\pi^2},
\end{multline}
where the effective electric and magnetic fields are 
\begin{equation}\label{effective}
\vec{E}_{eff}=\frac{d(\vec{A}+\vec{K})}{dt}+\nabla\mathcal E,\\ \ \ 
 \vec{B}_{eff}=\nabla\times(\vec{A}+\vec{K}) ,
 \end{equation}
 and $J=det(e)$ is the Jacobian; its sign corresponds to the chirality of the Weyl cone.

It is clear from ~\eqref{eq:topochar} that the anomaly equation is independent of dreibein and tilt vector -- it 
depends only on the existence of the Weyl point, 
 due to its topological nature.

Although the anomaly is not affected by dreibein and tilt vector, the general expression for the current still depends on them: 
\begin{equation}
\label{eq:cur}
\begin{split}
& j^i=\int \frac{d^3k}{(2\pi)^3} \sqrt{G}\dot{x^i}f=\int sgn(J)J^{-1} \frac{d^3p}{(2\pi)^3} \sqrt{G}\dot{x^i}f\\
&=\int sgn(J)J^{-1}  [-\Omega_{k_i t}+(\delta^i_j(1+\Omega_{k_lx^l})-\Omega_{k_ix^j})\pt_{k_j}\epsilon\\
&\qquad\qquad\qquad+\Omega_{k_ik_j}\pt_{x^j}\epsilon]f\frac{d^3p}{(2\pi)^3}.
 \end{split}
\end{equation}
The first term $\Omega_{k_it}$ describes a transverse current analogous to the anomalous Hall effect. The second term $\pt_{k_j}\epsilon$  corresponds to the anomalous velocity induced by lattice motion. To study the CME, we focus on the last three terms:
\begin{eqnarray}
\label{eq:cur1}
&\Omega_{k_jx^j}\pt_{k_i}\epsilon=\varepsilon_{abc}(e_a^j\Delta_{bj}e_d^i\frac{p_cp_d}{2|p|^4}+e_a^j\Delta_{bj}\frac{W_ip_c}{2|p|^3}),\\
\label{eq:cur2}
&\Omega_{k_ix^j}\pt_{k_j}\epsilon= \varepsilon_{abc}(e_a^i\Delta_{bj}e_d^j\frac{p_cp_d}{2|p|^4}+e_a^i\Delta_{bj}\frac{W_jp_c}{2|p|^3}),\\
 \label{eq:cur3}
&\Omega_{k_ik_j}\pt_{x^j}\epsilon= \varepsilon_{abc}(e_a^ie_b^j\Delta_{dj}\frac{p_cp_d}{2|p|^4}+e_a^ie_b^j\frac{\Gamma_j p_c}{2|p|^3}), 
\end{eqnarray}
where $\Delta_{ai}=\frac{\pt p_a}{\pt x^i}=\omega^j_{ai}T_j-e^j_a\frac{\pt A_i}{\pt x^j}$.
$\Gamma_{i}=\frac{\pt W^j}{\pt x^i}T_j-W^j\frac{\pt A_j}{\pt x^i}+\frac{\pt\mathcal E}{\pt x^i}$, and we have defined $\omega^i_{aj}=\frac{\pt e^i_a}{\pt x^j}$.

Using the symmetry relation $\epsilon_{abc}p_d- \epsilon_{dbc}p_a-\epsilon_{adc}p_b-\epsilon_{abd}p_c=0$ one can simplify these equations, and get the following expression for the anomalous CME current:
\begin{multline}
\label{eq:simp} 
j^i=\int sgn(J)J^{-1} \frac{d^3p}{(2\pi)^3}\times\\
[\Omega_{k_jx^j}\pt_{k_i}\epsilon-\Omega_{k_ix^j}\pt_{k_j}\epsilon
+\Omega_{k_ik_j}\pt_{x^j}\epsilon]f\\
=\int sgn(J)J^{-1} \frac{d^3p}{(2\pi)^3}\left[e_c^ie_b^j\omega_{na}^j\frac{T_a\epsilon_{bcn}}{2|p|^2}+det(e)\frac{B_{eff}^i}{2|p|^2}\right.\\
+\varepsilon_{abc}(e_a^k\omega^j_{bk}T_j W^i-e_a^i\omega^j_{bk}T_jW^k +e_a^i e_b^j\Gamma_j)\frac{p_c}{2|p|^3}\\
\left.+\frac{1}{2}\varepsilon_{abc}\varepsilon_{jlm}e_a^m e_b^l W^j\frac{p_c}{2|p|^3}B^i_{eff}\right]f
\end{multline}
 It is easy to derive from eq\eqref{eq:Ham} that $W^i=(0,-2t_3s_x\frac{\b_2}{t_2}\g zsin(k_{0}a),2t_3s_x\frac{\b_2}{t_2}\g ysin(k_{0}a))$ and $\gamma y, \gamma z \ll 1$, we will ignore it in our calculations, and treat the cones as untilted. In such a cone, in a equilibrium distribution, $f(|p|)$ can be assumed to be an even function of $p_a$. So, most terms in eq~\eqref{eq:simp} should vanish except the second term. It is easy to show  that the second term is also independent of dreibein. Therefore the CME current is topologically protected too. The contribution of each cone is
\begin{multline}
\label{eq:CME} 
j_n^i=\int sgn(J)J^{-1} \frac{d^3p}{(2\pi)^3}det(e)\frac{B^i_{eff}}{2|p|^2}f\\
=\chi_n\frac{(\m-\mathcal E_n) B_{eff}^i}{4\pi^2}
\end{multline}
where $\chi_n, \mathcal E_n$ are the chirality and energy of the $n_{th}$ Weyl point. In this expression, as explained in \cite{Basar:2013iaa}, 
both $\m$ and $\mathcal E_n$ should be counted from the bottom of the filled band. 

The evolution of the chiral charge can be determined in the relaxation time approximation:
\begin{equation}
\begin{split}
&\pt_i j^i_n + \dot{\rho}_n = \chi_n\frac{e^3\vec{E}\cdot \vec{B}}{4\pi^2}-\frac{e}{\t}(\rho_n-(\rho_n)_{eq})
\label{eq:uncon}
\end{split}
\end{equation} 
where $(\rho_n)_{eq}$ denotes the equilibrium chiral charge density in position space, and $\tau$ is the chirality relaxation time. So the chemical potential $\mu_n$ associated with the chiral charge is \cite{fukushima2008chiral,grushin2016inhomogeneous} 
\begin{equation}
\label{eq:ccp} 
\begin{split}
\delta \m_n = &\left(\frac{(\mu -\mathcal E_n)^2}{2\pi^2 v^x v^y v^z}\right)^{-1} \delta \rho_n\\ &= \chi \frac{16 \pi^2 ab^2 \cos(k_{x0}a) t_1 t_2^2}{(\mu-\mathcal E_n)^2}\frac{\vec{E}_{eff}\cdot\vec{B}_{eff}}{4\pi^2}\t
\end{split}
\end{equation}
where $\mu$ is the chemical potential. The contribution of each cone to the chiral magnetic current is given by 
\begin{equation}
\vec{j}_n =\chi_n \frac{e^2}{4\pi^2}\delta \mu_n \vec{B};
\end{equation}
therefore, the total CME current is determined by the chiral imbalance \cite{fukushima2008chiral}. 
The general form of the anomalous CME current in a deformed Weyl semimetal is given by ~\eqref{eq:simp}. 

\medskip
{\it The current in the model with torsion.---}
Following the general results derived above, we will now focus on the topology effect by elastic torsion--- effective gauge field induced on the Weyl points by torsion. Since the effective electric field transforms as a 2-form under arbitrary diffeomorphisms, the effective electric and magnetic fields (\ref{effective}) are:
\begin{equation}
\vec{E}_{eff} = s_x \frac{\b_1}{t_1 a}\cot(k_{0x}a)\dot{\lambda} \hat{x} ;
\end{equation}
 \begin{equation}
\vec{B}_{eff} = -s_x (s_y + s_z) \frac{\beta_2 \gamma  }{t_2 b} \sin(k_{0x} a)\hat{x} .
\end{equation}


The chiral anomaly at each cone is
\begin{equation}
\begin{split}
\vec{E}_{eff}\cdot \vec{B}_{eff}= 
-(s_y + s_z)\frac{\b_1 \b_2\ \cos(ak_{0x})\g\dot{\lambda} }{abt_1 t_2},
\end{split}
\label{eb}
\end{equation}



The total charge is conserved, as expected:
\begin{equation}
\begin{split}
\sum_n (\pt_i j^i + \dot{\rho})_n=&\sum_n\frac{\chi_n}{4\pi^2}(E_{eff}\cdot B_{eff})_n\\=& \sum\limits_{s_{x,y,z}=\pm 1} -s_y s_z (s_y + s_z)\frac{\b_1 \b_2\ \cos(ak_{0x})\g\dot{\lambda} }{abt_1 t_2}\\ = & 0 .
\end{split}
\end{equation}

\begin{figure}[hbtp]
\centering
\includegraphics[scale=0.6]{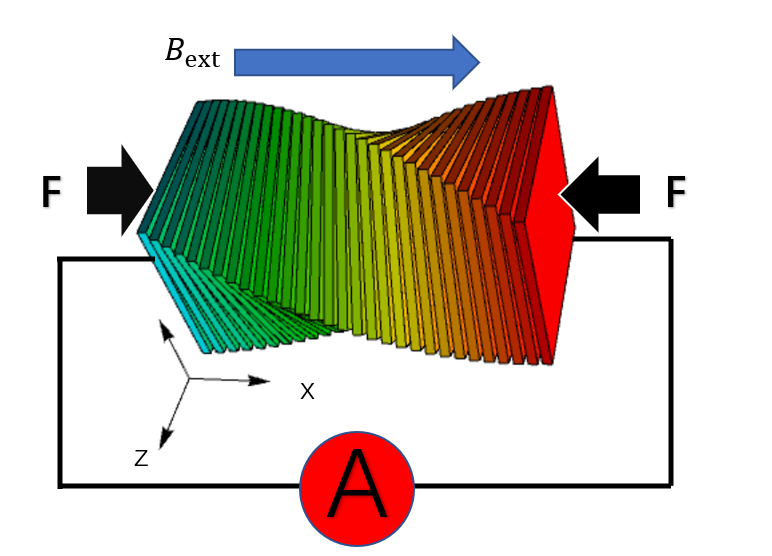}
\caption{An illustration of how the chiral imbalance and anomalous transport can be manipulated by deformations in an external magnetic field. The semimetal is twisted and compressed by the force F; this generates an electric current that is measured by an ammeter $A$.}
\label{fig:arch}
\end{figure}

The energies of the Weyl points are 
$$\mathcal E_n = t_3 (s_y + s_z); $$ therefore, 
combining equation~\eqref{eq:CME} and~\eqref{eq:ccp} we get
\begin{multline}
j_{total}^i = \sum_n(j^i)_n= \sum_n\frac{\chi_n \delta \m_n e^2 B_{total}^i}{4\pi^2}\\
=-4e^2\frac{\b_1\b_2 b t_2  \cos(ak_{0x})\g \dot{\lambda}}{\pi^2\hbar^2}\times\\
\left(\frac{1}{(\m-4t_3)^2}-\frac{1}{(\m+4t_3)^2}\right)\t B_{ext}^i.
\end{multline}
This shows that the current is linear in the external magnetic field \cite{fukushima2008chiral}, which is different from the CME induced in parallel (weak)  electric and magnetic fields, where $j\propto B^2$ \cite{son2013chiral,burkov2014chiral}. 
\medskip

For a numerical estimate, let us assume a statically twisted and dynamically compressed (along the axis of twist)  
crystal with a square cross section 2 mm x 2 mm, and take parameters $t_1 = 2 eV,  t_2 = 0.5 eV,  t_3 = 0.05 eV,  \b_1 = 1.5 eV,  \b_2 = 1 eV,  a = 0.3 nm,  b = 0.4 nm,  \tau = 10^{-10} s,  \mu = 0.15 eV,  ak_{0x} = \pi/3$. Let us also assume an external magnetic field of 10 T, with a twist parameter of $\g = 10 m^{-1}$ and compression rate $\dot{\lambda} = 10 s^{-1}$. The anomalous current density would be $\sim 470\ \mu A/m^2$, and the total current about 2 nA. Note that the current has an inverse square dependence on $\mu - \mathcal E_n$; it will be maximized in materials with a Fermi surface very close to a Weyl point. For example, for our parameters, if we take $\mu - 4t_3$ to be 0.01 eV instead of 0.05 eV, our current will increase by a factor of 25 to approximately 50 nA. 

\medskip

\section {Summary and outlook.}
Our generalized chiral kinetic equations (\ref{eq:kineq4}), (\ref{eq:kineq2}) and the anomaly equation (\ref{eq:topochar}) apply to any Weyl system with the Hamiltonian of the form $\sigma\cdot p(k,x,t)+\phi(k,x,t)$.  In the anomaly equation (\ref{eq:topochar}), torsion 
creates a synthetic magnetic field, while the time-dependent compressive strain creates a synthetic electric field -- so no external electric field is necessary to generate the chiral chemical potential. Detecting the resulting torsion-induced chiral magnetic current would thus allow to establish the anomalous coupling between the spatial and momentum-space chiralities,  without a background from the Ohmic current that exists in the longitudinal magnetoresistance measurements. 

The current measured in the corresponding experiment will also not get contributions from the piezoelectric effect (because the material is inversion symmetric and has no piezoelectricity), or from eddy currents (because the compression is along the magnetic field). 

It has been suggested that lattice defects can act as sources of emergent gravity: a disclination induces a curvature, and a dislocation induces torsion\cite{imaki2019lattice}. It would be interesting to extend our method to computing the anomalous currents resulting from the combination of chiral and gauge-gravitational anomalies in this case.

\begin{acknowledgments}
We thank Xuguang Huang, Jennifer Cano, Mengkun Liu and Qiang Li for useful discussions. This
work was supported in part by the China Scholarship Council (L-L. G.), U. S. Department of
Energy under Awards DE-SC0017662 (S. K. and D. E.
K.),
 DE-FG02-88ER40388 (E. J. P. and D. E. K.) and DE-SC0012704 (D. E. K.).

\end{acknowledgments}

\appendix
\section{The detailed derivation of eq\eqref{eq:kineq4}-\eqref{eq:kineq3}}
From the Euler-Lagrange equation for the momenta we get:
\ba
\frac{\pt I}{\pt k_i}-\frac{d\pt I}{dt\pt \dot{k}_i}=0 
\ea
we derive
\ba
\dot{x}_i=\pt_{k_i}\epsilon -\Omega_{kixj}\dot{x}_j-\Omega_{kikj}\dot{k}_j .
\ea

The Euler-Lagrange equation for the coordinates yields
\ba
\frac{\pt I}{\pt x_i}-\frac{d\pt I}{dt\pt \dot{x}_i}=0 
\ea
from which we derive
\ba
\dot{k}_i=-\pt_{x_i}\epsilon +\Omega_{xikj}\dot{k}_j+\Omega_{xixj}\dot{x}_j .
\ea
Since these two equations are coupled, the solution can be found using a method based on differential geometry. This work has already been done in \cite{Hayata:2016wgy}  by T. Hayata and Y. Hidaka. Their results are:
\ba
\label{A5}
\sqrt{G}&=&1+\Omega_{k_ix_i}-\Omega_{x_ix_j} \Omega_{k_ik_j}+((\Omega_{k_ix_i}) ^2-\Omega_{k_ix_j} \Omega_{k_jx_i} )/2\non
&&-\epsilon_{ikl}\epsilon_{jmn} \Omega_{k_kk_l}\Omega_{k_ix_j} \Omega_{x_mx_n}/4\non
&&+ \epsilon_{ikl}\epsilon_{jmn}\Omega_{k_jx_i} \Omega_{k_mk_k}\Omega_{k_nx_l} /6
\ea
\ba 
\label{A6}
\sqrt{G}\dot{x}&=&(\delta_{ij}(1+\Omega_{k_kx_k})-\Omega_{k_ix_j}-\epsilon_{ikl}\epsilon_{jmn}\Omega_{k_mk_n}\Omega_{x_kx_l}/4\non
&&+\epsilon_{ikl}\epsilon_{jmn}\Omega_{k_mx_k} \Omega_{k_nx_l}/2)(\pt_{k_j}\epsilon-\Omega_{k_jt})\non
&&+(\Omega_{k_ik_j}+\epsilon_{kls}\epsilon_{ijm}\Omega_{k_kx_m} \Omega_{k_lk_s}/2)(\pt_{x_j}\epsilon-\Omega_{x_jt})\non
\ea
\ba 
\label{A7}
\sqrt{G}\dot{k}&=&\delta_{ij}(1+\Omega_{k_kx_k})-\Omega_{k_jx_i}-\epsilon_{ikl}\epsilon_{jmn}\Omega_{k_mk_n}\Omega_{x_kx_l}/4\non
&&+\epsilon_{ikl}\epsilon_{jmn}\Omega_{k_mx_k} \Omega_{k_nx_l}/2)(-\pt_{x_j}\epsilon +\Omega_{x_jt})\non
&&-(\Omega_{x_ix_j}+\epsilon_{kls}\epsilon_{ijm}\Omega_{k_kx_m} \Omega_{k_lk_s}/2)(\pt_{k_j}\epsilon-\Omega_{k_j t})\non
\ea

Here we will consider the case of a Weyl-type Hamiltonian:
$H=\sigma\cdot p+\phi$. The equation \eqref{eq:Weq-full} is a key point for deriving \eqref{eq:kineq4}, 
\eqref{eq:kineq2}, and \eqref{eq:kineq3}. 

Since \begin{equation}
\begin{split}
   &\Omega_{\alpha\beta}=\partial_b a_\alpha-\partial_a a_\beta=\frac{\partial p_j}{\partial \beta}\frac{\partial p_i}{\partial \alpha}\frac{\partial a_p^i}{\partial p_j}-\frac{\partial p_j}{\partial \alpha}\frac{\partial p_i}{\partial \beta}\frac{\partial a_p^i}{\partial p_j}\\
   &=\frac{1}{2}\partial_\alpha p_m \partial_\beta p_l\epsilon_{mln}\Omega_n^p ,
   \end{split}
\end{equation}
we have the following relation:
\begin{equation}
\begin{split}
 & \Omega_{\alpha\beta} \Omega_{\gamma\sigma}\\ =&\varepsilon_{mnl}\varepsilon_{m'n'l'}\frac{\partial p_m}{\partial \alpha}\frac{\partial p_n}{\partial \beta}\frac{\partial p_m'}{\partial \gamma}\frac{\partial p_n'}{\partial \sigma}\Omega^p_l\Omega^p_{l'}\\
&=(\frac{\partial p \cdot \partial p}{\partial \alpha \partial\gamma}\frac{\partial p \cdot \pt p}{\partial \beta \partial\sigma}-\frac{\partial p \cdot \pt p}{\partial \alpha \partial\sigma}\frac{\partial p \cdot \partial p}{\partial \beta \partial\gamma})\Omega^2\\
&-\frac{\Omega \cdot \partial p}{\partial \alpha }(\frac{\Omega \cdot \partial p}{\partial \gamma }\frac{\partial p \cdot \partial p}{\partial \beta \partial\sigma}-\frac{\Omega \cdot \partial p}{\partial \sigma }\frac{\partial p \cdot \partial p}{\partial \beta \partial\gamma})\\
&+\frac{\Omega \cdot \partial p}{\partial \beta }(\frac{\Omega \cdot \partial p}{\partial \gamma }\frac{\partial p \cdot \partial p}{\partial \alpha \partial\sigma}-\frac{\Omega \cdot \partial p}{\partial \sigma }\frac{\partial p \cdot \partial p}{\partial \alpha \partial\gamma})
\end{split}
\end{equation}

Therefore the following cancellation takes place, as shown in \eqref{eq:Weq-full}:
\begin{equation}
    \begin{split}
 &\Omega_{\alpha\beta}\Omega_{\gamma\sigma}+\Omega_{\alpha\gamma}\Omega_{\sigma\beta}+\Omega_{\alpha\sigma}\Omega_{\beta\gamma}\\
&=(\frac{\partial p \cdot \partial p}{\partial \alpha \partial\gamma}\frac{\partial p \cdot \partial p}{\partial \beta \partial\sigma}-\frac{\partial p \cdot\partial p}{\partial\alpha \partial\sigma}\frac{\partial p \cdot \partial p}{\partial \beta \partial\gamma})\Omega^2\\
&+(\frac{\partial p \cdot \partial p}{\partial \alpha\partial\sigma}\frac{\partial p \cdot \partial p}{\partial \gamma \partial\beta}-\frac{\partial p \cdot \partial p}{\partial \alpha \partial\beta}\frac{\partial p \cdot \partial p}{\partial \sigma \partial\gamma})\Omega^2\\
&+(\frac{\partial p \cdot \partial p}{\partial \alpha \partial\beta}\frac{\partial p \cdot \partial p}{\partial \sigma \partial\gamma}-\frac{\partial p \cdot \partial p}{\partial\alpha \partial\gamma}\frac{\partial p \cdot \partial p}{\partial \sigma \partial\beta})\Omega^2\\
&-\frac{\Omega \cdot \partial p}{\partial \alpha }(\frac{\Omega \cdot \partial p}{\partial \gamma }\frac{\partial p \cdot \partial p}{\partial \beta \partial\sigma}-\frac{\Omega \cdot \partial p}{\partial \sigma }\frac{\partial p \cdot \partial p}{\partial \beta \partial\gamma}\\
&+\frac{\Omega \cdot \partial p}{\partial \sigma }\frac{\partial p \cdot \partial p}{\partial \beta \partial\gamma}-\frac{\Omega \cdot \partial p}{\partial \beta }\frac{\partial p \cdot \partial p}{\partial \gamma \partial\sigma}\\
&+\frac{\Omega \cdot \partial p}{\partial \beta}\frac{\partial p \cdot \partial p}{\partial \gamma \partial\sigma}-\frac{\Omega \cdot \partial p}{\partial \gamma }\frac{\partial p \cdot \partial p}{\partial \sigma\partial\beta})\\
&+\frac{\partial p}{\partial \alpha}\cdot(\frac{\partial p}{\partial \sigma}\frac{\Omega \cdot\partial p}{\partial \beta }\frac{\Omega \cdot \partial p}{\partial\gamma }-\frac{\partial p}{\partial \gamma}\frac{\Omega \cdot \partial p}{\partial \sigma }\frac{\Omega \cdot \partial p}{\partial \beta }\\
&+\frac{\partial p}{\partial \gamma}\frac{\Omega \cdot \partial p}{\partial \beta }\frac{\Omega \cdot \partial p}{\partial \sigma }-\frac{\partial p}{\partial \sigma}\frac{\Omega \cdot \partial p}{\partial \beta }\frac{\Omega \cdot \partial p}{\partial \gamma}\\
&+\frac{\partial p}{\partial \sigma}\frac{\Omega \cdot \partial p}{\partial \gamma}\frac{\Omega \cdot \partial p}{\partial \beta }-\frac{\partial p}{\partial \beta}\frac{\Omega \cdot \partial p}{\partial \sigma }\frac{\Omega \cdot \partial p}{\partial \gamma })\\
&=0
\end{split}
\end{equation}
 The \eqref{eq:Weq-full} makes the following terms vanish:
\ba 
(\epsilon_{ikl}\epsilon_{jmn}\Omega_{k_mx_k} \Omega_{k_nx_l}/2 -\epsilon_{ikl}\epsilon_{jmn}\Omega_{k_mk_n}\Omega_{x_kx_l}/4)=0\nonumber
\ea
\ba
\epsilon_{kls}\epsilon_{ijm}\Omega_{k_kx_m} \Omega_{k_lk_s}/2=0 
\ea
\ba
 -(\Omega_{k_kx_k}-\Omega_{k_ix_j})\Omega_{k_jt}-\Omega_{k_ik_j}\Omega_{x_jt}=0
\ea

 Finally the equations \eqref{A5}-\eqref{A7} can be simplified to
\ba
&&\sqrt{G}=1+\Omega_{k_ix_i}\\
 &&\sqrt{G}\dot{x}=-\Omega_{k_i t}+(\delta_{ij}(1+\Omega_{k_kx_k})-\Omega_{k_ix_j})\pt_{k_j}\epsilon+\Omega_{k_ik_j}\pt_{x_j}\epsilon\non
\ea
\ba
 \sqrt{G}\dot{k}=\Omega_{x_i t}-(\delta_{ij}(1+\Omega_{k_kx_k})-\Omega_{k_ix_j})\pt_{x_j}\epsilon-\Omega_{x_ix_j}\pt_{k_j}\epsilon\non
\ea

The Liouville anomaly equation is easily derived from above three equation:
\ba
  \frac{\pt\sqrt{G}}{\pt t}+\frac{\pt\sqrt{G}\dot{x}_i}{\pt x_i}+\frac{\pt\sqrt{G}\dot{k}_i}{\pt k_i}&=&\Theta_{x_i k_i t}+\Theta_{k_jx_jx_i}\pt_{k_i}\epsilon\non
  &&-\Theta_{x_jk_jk_i}\pt_{x_i}\epsilon
\ea

\section{The summation over cones}
The chirality and effective gauge fields depend on $s_x,s_y$ and $s_z$ as we explained below eq(7).
To determine the total anomalous current and the total chiral charge generation, we should take into account the following summation rule:
\begin{eqnarray}
&\sum_n s_i=0(i=x,y,z)\\
&\sum_n s_is_j=0(i\neq j)\\
& s_is_i=1
\end{eqnarray}
Using this summation rule, we find that the current $j_{tor}$ originating from torsion alone and the total current $j_{tot}$ resulting from the combination of the torsional electric field and an external magnetic field vanish after summation over all cones:
\begin{equation}
\begin{split}
&j_{tor}\propto \sum_n \frac{1}{(\m-\mathcal E_n)^2}( E_{eff}\cdot B_{eff})B_{eff}\\
&=\sum_n \frac{s_x}{(\m-(s_y+s_z)t_3)^2}  \frac{\b_1}{t_1 a}\cot(k_{0x}a)\dot{\lambda}|B_{eff}|^2\\
&=\sum_n \frac{s_x(s_y+s_z)^2}{(\m-(s_y+s_z)t_3)^2}  \frac{\b_1}{t_1 a}\cot(k_{0x}a)\dot{\lambda}(\frac{\b_2\gamma}{t_2b})^2sin^2(k_{0x}a)\\
&=0 .
\end{split}
\end{equation}
\begin{equation}
\begin{split}
&j_{tot}\propto \sum_n \frac{1}{(\m-\mathcal E_n)^2}( E_{eff}\cdot B_{ext})B_{ext}\\
&=\sum_n \frac{s_x}{(\m-(s_y+s_z)t_3)^2}  \frac{\b_1}{t_1 a}\cot(k_{0x}a)\dot{\lambda}B_x \vec B=0 .
\end{split}
\end{equation}
Therefore the only current that survives the summation over the cones is the chiral magnetic current resulting from the chiral chemical potential induced by strain in an external magnetic field.

\section{Cutoff independence of the monopole charge functions}

In this section we show that the monopole charge function computed from the entire tight-binding Hamiltonian \eqref{eq:Ham}coincides with the one computed for an effective low energy Weyl Hamiltonian. 

 From eq\eqref{eq:Ham} we get:
\ba
\label{m}
&p_1=-2t_1(cos(k'_xa)-cos(k'_{x0}a))+2\b_1 \l cos(k'_xa)\\
\label{m1}
&p_2=-2t_2sin(k'_yb)+2\b_2\g zsin(k'_xa)\\
\label{m2}
&p_3=-2t_2sin(k'_zb)-2\b_2\g ysin(k'_xa)\\
\label{m3}
&\phi=2[cos(k'_yb)+cos(k'_zb)]t_3
\ea
where $k_i'=M_i^j k_j$, 
so the dependence of the $p_a$ on the variables is given by:

 $p_1(k_x,\l(t))$,
 
  $p_2(k_x,k_y,k_z,x,z,\l(t))$,
  
   $p_3(k_x,k_y,k_z,x,y,\l(t))$

Let us use the above equations to compute the monopole charge functions responsible for the chiral anomaly:
\ba
\Theta_{k_ix^it}&=&\varepsilon_{mnl}\frac{\pt p_m}{\pt k_i} \frac{\pt p_n}{\pt x^i}\frac{\pt p_l}{\pt t}  \Theta(p)\non
&=& \Big[-\frac{\pt p_1}{\pt \l}\dot \l(\frac{\pt p_2}{\pt z}\frac{\pt p_3}{\pt k_z}+\frac{\pt p_2}{\pt x}\frac{\pt p_3}{\pt k_x})\non
&&+\frac{\pt p_1}{\pt \l}\dot \l(\frac{\pt p_3}{\pt y}\frac{\pt p_2}{\pt k_y}+\frac{\pt p_3}{\pt x}\frac{\pt p_2}{\pt k_x})\non
&&+ \frac{\pt p_2}{\pt \l}\dot \l(\frac{\pt p_1}{\pt x}\frac{\pt p_3}{\pt k_x}-\frac{\pt p_3}{\pt x}\frac{\pt p_1}{\pt k_x})
\non
&&-\frac{\pt p_3}{\pt \l}\dot \l(\frac{\pt p_1}{\pt x}\frac{\pt p_2}{\pt k_x}-\frac{\pt p_2}{\pt x}\frac{\pt p_1}{\pt k_x})\Big]\Theta(p)\non
&=&C\Theta(p)
\ea
where, for simplicity,  we use $C$ to denote the part inside the square bracket. $C$ contains four terms separately displayed by the four line above. For the first term, which we denote as $C_1$ we have
\ba
C_1&=&-\frac{\pt p_1}{\pt \l}\dot \l(\frac{\pt p_2}{\pt z}\frac{\pt p_3}{\pt k_z}+\frac{\pt p_2}{\pt x}\frac{\pt p_3}{\pt k_x})\non
&=&-2\b_1cos(k'_xa)\dot \l(-4t_2\b_2\g sin(k'_xa)bcos(k'_zb)\non
&&+4t_2\b_2 b a
cos(k'_y b)\frac{dk'_y}{dx}\g ycos(k'_x a)\frac{dk'_x}{dk_x})\non
&=&-2\b_1cos(k'_xa)\dot \l(-4t_2\b_2\g sin(k'_xa)bcos(k'_zb)\non
&&+4t_2\b_2 b a
cos(k'_y b)\g k_z\g ycos(k'_x a)(1-\l))\nonumber
\ea

Since $\Theta_{kxt}= C\Theta(p)$, where $\Theta(p)=2\pi\d^3(p)$, we are only interested in the value of $C_1$ at the point $p=0$:
\ba
C_1|_{p=0}&=&C_1|_{k=K}=2\b_1cos(K_{0x}a)\dot \l\non
&&\times(-4t_2\b_2\g sin(K_{0x}a)bs_z+4t_2\b_2 b a
\g K_z\g y(1-\l))\non
&=&-8\b_1cos(K_{0x}a)\dot \l t_2\b_2\g sin(K_{0x}a)bs_z+0(\g^{2})\nonumber
\ea

The same way, one can derive the second term of $C$ to be:
\ba 
C_2|_{k=K}&=&2\b_1cos(K_{0x}a)\dot \l\non
&&\times(-4t_2\b_2\g sin(K_{0x}a)bs_y-4t_2\b_2 b a
\g K_y\g z(1-\l))\non
&=&-8\b_1cos(K_{0x}a)\dot \l t_2\b_2\g sin(K_{0x}a)bs_y+0(\g^{2})\nonumber
\ea

The third term is of higher order in deformation:
\ba
C_3&=&4\b_1\g zcos(k'_x a)a\frac{dk'_x}{d\l}\dot \l t_2cos(k'_zb)b\frac{dk'_z}{dx}\non
&&\cdot (2t_1-2\b_1\l)sin(k'_xa)a(1-\l)\non
C_3|_{k=K}&=&-8\b_1\g zcos(K'_{x0} a)aK_x\dot \l t_2b\g K_y\non
&&\cdot (t_1-\b_1\l)sin(K'_xa)a(1-\l)=0(\g^{2})\nonumber
\ea
Similarly, the last term $C_4$ is also a higher order correction.

So, the integration of monopole charge function yields
\ba
\int \frac{d^3k}{2\pi}\Theta_{k_ix^it}&=& \int \frac{d^3p}{(2\pi)^3}[J^{-1}C]\cdot 2\pi\d^3(p)\non
&=&\frac{1}{4\pi^2}[det(J)^{-1}C]|_{k_i=K_i}\label{en}
\ea
where $J$ is a Jacobian. If we define $e_a^i=\frac{d p_a}{d k_i}$ , $J=det(e)$. Since $e_a^i=(e')_a^j M_j^i$ , 
\ba
 det(J)^{-1}C|_{k_i=K_i}=det(e')^{-1}det(M)^{-1}C|_{k_i=K_i}\label{em}
\ea

From\eqref{m}-\eqref{m3} :
\ba
&&(e')_a^i=\frac{dp_a}{dk'_j}\non
&&={\left[ \begin{array}{ccc}
(2t_1 -2\b_1\l)sin(k'_x a)a& 0& 0 \\
2\b_2 \g zsin(k'_x a)a&-2t_2cos(k'_y b)b &0\\
-2\b_2 \g ysin(k'_x a)a&0 &-2t_2cos(k'_y b)b
\end{array}
\right]}
\nonumber
\ea

To count the lowest order, we should take the lowest order of $det(e')^{-1}|_{k_i=K_i}$,$det(M)^{-1}|_{k_i=K_i}$ and $C|_{k_i=K_i}$ separately as below:
\ba\label{q0}
&&det(e')^{-1}|_{k_i=K_i}=\frac{1}{8t_1t_2^2sin(K'_xa)ab^2} \\
\label{q1}
&&det(M)^{-1}=1\\
\label{q2}
&&C=C_1+C_2+C_3+C_4\non
&=&-8(s_y+s_z)\b_1\b_2\g\dot \l t_2 absin(K_{0x}a)cos(K'_{0x}a)
\label{q3}
\ea

One can also prove that in the first order of $\g$  we get
\ba
\Theta_{k_jx_jx_i}\pt_{k_i}\epsilon+\Theta_{x_jk_jk_i}\pt_{x_i}\epsilon = 0
\ea

So, finally let us multiply \eqref{q0}, \eqref{q1}, and \eqref{q2} to get the result for the anomaly:
\ba
\pt_\m j^\m&=&\int \frac{d^3k}{2\pi}\Theta_{k_ix^it}=\frac{1}{4\pi^2}[det(J)^{-1}C]|_{k_i=K_i}\non
&=&-\frac{cos(K'_{0x}a)\b_1\b_2\g\dot \l}{4\pi^2 t_1 t_2ab} (s_y+s_z)\label{el}
\ea
It coincides with eq\eqref{eb} in our manuscript, which means that the computation with the effective Hamiltonian does not miss any anomaly terms in lowest order in deformation.  

This means that we do not need to impose a cut-off. In principle, we can go beyond the lowest order using the entire Hamiltonian of eq\eqref{eq:Ham}, but the result becomes cumbersome at higher order. 
We also do not need a cut-off in computing the monopole charge function in the case of elastic continuous deformation.


\bibliography{bibfile}

\end{document}